# Light–matter interaction in van der Waals heterostructures with Mie voids

Zhuoyuan Lu[1,3], Kirill Koshelev[2*], Pavel Tonkaev[4], Ziyu Chen[1], Dawei Liu[2], Wenkai Yang[1], Yuri Kivshar[4*] and Yuerui Lu[1,3*]

[1] School of Engineering, College of Engineering, Computing and Cybernetics, Australian National University, Canberra ACT 2601, Australia

[2] Department of Electronic Materials Engineering, Research School of Physics, Australian National University, Canberra ACT 2601, Australia

[3] ARC Centre for Quantum Computation and Communication Technology, Australian National University, Canberra ACT 2601, Australia

[4] Nonlinear Physics Center, Research School of Physics, Australian National University, Canberra ACT 2601, Australia

* To whom correspondence should be addressed:
Kirill Koshelev (kirill.koshelev@anu.edu.au), Yuri Kivshar (yuri.kivshar@anu.edu.au) and Yuerui Lu (yuerui.lu@anu.edu.au)

**Abstract:**

Recently introduced concept of Mie voids allows to enhance the field localization inside air cavities embedded in high-index materials. Mie voids provide an alternative approach to conventional dielectric resonators that confine optical fields within bulk high-index materials. Building on this concept, here we present a hybrid photonic platform that integrates monolayer $WS_2$ with Mie void resonators patterned in a high-index $Bi_2Te_3$ substrate. By carefully aligning the dipolar void resonance with the excitonic transition of $WS_2$, we achieve substantially enhanced photoluminescence and second-harmonic generation. Far-field imaging of the harmonic fields reveals spatially resolved hotspots that directly map localized resonant modes, with their positions tunable by cavity geometry and pump wavelength. This approach enables real-space control of nonlinear emission at the single-resonator level, offering a robust and reconfigurable platform for next-generation nonlinear photonics and surface-enhanced optical sensing.

# Introduction

Controlling light–matter interaction at the nanoscale is central to emerging nonlinear and quantum photonic technologies. Conventional dielectric resonators confine optical fields within bulk high-index materials, limiting tunability and surface accessibility. Recently introduced concept of Mie voids offer an alternative approach to enhance the field localization near interfaces and reduce sensitivity to fabrication imperfections[1,2]. In general, subwavelength photonic resonators enable fine control over light–matter interaction at the nanoscale, allowing spectral shaping, field enhancement, and tailored multipolar responses for applications in nonlinear optics, quantum emission, and sensing[3-6]. In this context, two-dimensional (2D) materials have emerged as a compelling class of ultrathin optical media. Materials such as transition metal dichalcogenides (TMDs) exhibit strong excitonic resonances[7], large oscillator strengths[8], and broken symmetry-enabled second-order nonlinearities[9-11]. Their atomic-scale thickness allows them to be transferred onto diverse photonic platforms without lattice-matching constraints, and has enabled demonstrations of photoluminescence (PL)[12], second-harmonic generation (SHG)[13], and spontaneous parametric down-conversion (SPDC)[14,15].

Despite these advantages, the intrinsically low optical thickness of 2D materials imposes fundamental constraints on their interaction volume. This limitation affects the strength of light–matter coupling, including SHG, exciton dynamics, field modulation, and directional scattering[16,17]. To address this, dielectric nanostructures and metasurfaces have been widely used to concentrate optical fields at the material interface, thereby amplifying emission and nonlinear processes[18-21]. Conventional dielectric metasurfaces typically rely on large-area periodic arrays of resonant units, where the optical response is governed by inter-element coupling and lattice symmetry. Such architectures offer limited spatial tunability and are highly sensitive to fabrication imperfections.

Mie void resonators — subwavelength air cavities carved into high-index materials — provide an alternative resonant mechanism with three key advantages. First, Mie voids confine resonant modes within low-index regions, where strong Fresnel reflection enables robust field localization without relying on total internal reflection. This unique confinement mechanism enhances tunability and facilitates efficient interaction with surface-bound materials. Second, the vertical position of the resonant field is tunable via cavity geometry, allowing precise alignment with overlaid 2D materials or other functional layers to optimize interaction. Third, void resonators remain effective in highly absorptive media, such as bismuth telluride ($Bi_2Te_3$), where conventional dielectric resonators would fail due to damping losses[22]. These features collectively enable a platform that is spatially localized, geometrically tunable, and compatible with materials that are challenging to use in standard photonic designs.

Here, we present a van der Waals hybrid photonic platform that integrates $Bi_2Te_3$-based Mie voids capped with monolayer $WS_2$, enabling spectrally and spatially tunable resonant light–matter interactions. $Bi_2Te_3$ features both a high refractive index and a large extinction coefficient, making it well suited for supporting void resonances with strong confinement[22]. By carefully designing cavities, we align the dipolar void surface resonances with the A-exciton energy of $WS_2$, resulting in strong enhancement of nonlinear optical processes[23]. We observe approximately 20x enhancement in PL and 25x enhancement in SHG from monolayer positioned on resonant Mie void compared to that on an off-resonant void.

Beyond linear and nonlinear enhancement, our single-resonator platform uniquely enables deterministic spatial control and real-space visualization of optical modes — a capability previously inaccessible in collective-resonator systems. Through hyperspectral imaging, we directly track mode profiles and systematically tune their resonances via cavity geometry. This approach allows exploration of localized field–exciton coupling, mode competition, and spatially dependent nonlinear processes. Rather than relying on extreme Q-factors or collective designs, we offer an experimentally flexible, geometrically tunable, and spatially resolved platform, opening new directions for nonlinear and spatially selective photonic applications.

## Results and Discussion

**Van der Waals heterostructures of $Bi_2Te_3$ Mie Voids with $WS_2$ monolayers**

Mie voids provide a unique photonic platform in which optical resonances are confined within air cavities embedded inside high-index dielectric hosts. Unlike conventional dielectric resonators, where the field is concentrated in the solid material, the Mie void geometry inverts this picture: resonant modes are localized in the low-index region due to strong Fresnel reflection at the air–dielectric interface[1]. This confinement mechanism enables high quality factors, strong field localization, and spectral tunability without requiring photonic bandgaps or plasmonic components. These features make Mie voids especially attractive for light-matter interaction.

Previous implementations have typically employed silicon or gallium arsenide as the host material, where the refractive index n lies in the range of 3–4, and the extinction coefficient *k* is nearly zero in the visible–near-infrared range[24]. In contrast, van der Waals $Bi_2Te_3$ exhibits an unusually high refractive index (n≈ 6–8) along with an extremely high extinction coefficient (k≈ 4–5) across the same spectral range[22]. While large k values usually lead to strong absorption and degraded resonance in conventional systems, Mie voids are uniquely tolerant of such absorption due to their air-core configuration minimizing field intensity inside the absorbing host and even leveraging the absorption to improve boundary reflectivity and modal confinement.

The resonant wavelengths of Mie modes of a cylinder-shaped void can be approximated with the solutions for a spherical void geometry because void modes are less sensitive to the shape effect compared to conventional dielectric nanoantennas[1,2]. The Mie solution for voids can be written explicitly in the limit of large refractive index contrast as[2]:

$$kr = m\pi + \arg(R) + \phi_m - i \ln\left(\frac{1}{|R|}\right), \quad (1)$$

where $m$ is the mode order, $k = (2\pi)/\lambda$ is the vacuum wavenumber, $R = (\sqrt{\varepsilon} - 1)/(\sqrt{\varepsilon} + 1)$ is the complex Fresnel reflection coefficient at the interface of the void and high-index host medium, $\varepsilon = (n + ik)^2$ is the complex dielectric constant of the host medium, $\arg(R)$ accounts for the phase shift induced by internal reflection, and $\phi_m$ describes small corrections of $\frac{1}{\sqrt{\varepsilon}}$ order that can vary for modes of different type. This expression highlights the essential role of cavity size parameter $\frac{r}{\lambda}$ and refractive index $\sqrt{\varepsilon}$ in determining the resonant frequencies and mode Q factors, and provides a useful model for analyzing design-dependent tuning and modal overlap[25]. We emphasize that the dielectric constant does not appear in the wavenumber in Eq. (1) as the void fields are confined within the air region. Deviation of geometry from spherical to cylindrical can be treated as a small correction to $\phi_m$ in Eq. (1) that depends on $\frac{d}{r}$, where $d$ and $r$ is the void depth and radius, respectively.

From Eq. (1), we can evaluate the mode quality ($Q$) factor as

$$Q = \frac{Re[k]}{-2Im[k]} = \frac{m\pi + \arg(R) + \phi_m}{2\ln\left(\frac{1}{|R|}\right)}. \quad (2)$$

We next analyze the dependence of Q on the dielectric constant for highly transparent and highly absorbing host material cases. In both cases, for $|\varepsilon| \to \infty$ the reflection becomes perfect, $|R| \to 1$, thus the mode becomes completely decoupled from the radiation continuum, $Q \to \infty$. For a highly transparent host medium $(n \gg k)$, we can estimate the mode losses as $\ln\left(\frac{1}{|R|}\right) \approx \frac{2}{n}$, thus the quality factor scales as $Q \propto \sqrt{|\varepsilon|}$. For the opposite case of highly absorbing host medium ($n \ll k$), we can estimate the mode losses as $\ln\left(\frac{1}{|R|}\right) \approx \frac{2n}{k^2}$, thus the quality factor scales as $Q \propto |\varepsilon|$. Combined, we can write that

$$Q \propto |\varepsilon|^\alpha, \tag{3}$$

where $\alpha$ is the power scaling exponent equal to $\alpha = \frac{1}{2}$ for $n \gg k$, and to $\alpha = 1$ for $n \ll k$.

To demonstrate the working principle, we propose a hybrid van der Waals heterostructure composed of vertically milled $Bi_2Te_3$ Mie voids capped by monolayer $WS_2$ (Fig. 1a). The air cavity supports surface-localized resonances, which couple directly to the excitonic transitions in the 2D material. To optimize the resonance, we carried out COMSOL simulations scanning over cavity radius and depth. As shown in Fig. 1b, under 625 nm excitation (matching the $WS_2$ A-exciton), the maximum in-plane electric field enhancement $|E|^2$ at the void top surface occurs for a radius of ~850 nm and depth of ~780 nm. This configuration supports the electric dipole mode and ensures optimal field–exciton overlap with the monolayer, while maintaining fabrication feasibility.

We further examined the spatial evolution of the resonance with cavity depth. Side-view simulations in Fig. 1c reveal that for shallow voids, the resonance remains near the surface, while deeper cavities exhibit a gradual downward shift in the mode profile. This spatial migration correlates with a spectral redshift and reflects the strong depth–wavelength–position coupling inherent to the void geometry, providing an accessible route for spatial and spectral mode engineering[26].

To evaluate the role of material properties in the enhancement behaviour, we performed a comparative analysis of $Bi_2Te_3$ and Si Mie voids under identical structural conditions. From the temporal coupled mode theory[27,28], it follows that:

$$|E|^2 \propto \kappa \frac{Q}{V}, \tag{4}$$

where $|E|^2$ is the peak surface electric field enhancement, $V$ is the mode volume, $\kappa$ is the coupling coefficient that characterized the spatial overlap between the pump and mode fields. We note that the absorption Q factor does not appear in Eq. (4), because the mode is localized within the air and the absorption only modifies the interface reflection coefficient, see Eq. (2). Combining Eq. (4) with Eq. (3) for Si and Bi$_2$Te$_3$, we can write

$$\frac{|E_{Bi2Te3}|^2}{|E_{Si}|^2} \propto \left|\frac{\varepsilon_{Bi2Te3}}{\varepsilon_{Si}}\right|^\alpha, \tag{5}$$

where the value of $\alpha$ can be estimated from Eq. (3) to be between ½ and 1. As shown in Fig. 1d, Bi$_2$Te$_3$ consistently yields twofold field enhancement than Si at the wavelengths 625 nm, 780 nm, and 1550 nm validating Eq. (5). We note that in contrast to conventional nanoantennas, high absorption losses in the void host material only increase |Ɛ| showing dominance of such materials over transparent dielectrics for Mie-void photonics.

Despite significant differences in dispersion between Bi$_2$Te$_3$ and Si, both materials support well-defined Mie resonances across the visible to near-infrared spectrum. While the resonance magnitude is clearly influenced by material dispersion — with stronger dispersion leading to higher field enhancement and increased quality factor — the existence and spectral position of the modes remain stable. This dependence of field intensity and Q-factor on dispersion is consistent with previous theoretical studies, which predict that stronger material dispersion improves confinement via enhanced Fresnel reflection. This behavior confirms that the resonance condition is primarily governed by geometry, enabling broadband and dispersion-resilient designs for nonlinear photonic applications[29], and is further confirmed by the results shown in Supplementary Fig.1,2.

**Resonant manipulation of Mie voids**

To experimentally validate the theoretical predictions established in Section 1, we fabricated high-quality Mie void arrays by focused ion beam (FIB) milling on mechanically exfoliated $Bi_2Te_3$ flakes with thicknesses exceeding 10 μm[30]. The resulting voids exhibit smooth sidewalls, high aspect ratios, and uniform spacing, as shown in Fig. 2a, ensuring optical symmetry and well-defined boundary conditions. All arrays were designed with sufficient edge-to-edge separation larger than half of the void radius to prevent coupling between adjacent cavities, guaranteeing that each resonator behaves as an isolated optical unit[31]. The negligible influence of inter-void coupling on the resonance behavior is further confirmed by numerical simulations presented in Supplementary Fig.2.

To determine the optimal design conditions for field enhancement, we performed numerical simulations by systematically sweeping the void depth and radius. The peak electric-field enhancement occurs at approximately a depth-to-radius ratio of $d/r \approx 0.94$. as further confirmed in Supplementary Fig.3. According to previous theoretical predictions[1], this condition corresponds to a dipolar-mode–dominated field enhancement, which is relatively insensitive to small fabrication imperfections and to variations in the material refractive index, and is thus well suited for experimental realization, as illustrated in Fig. 1b. Guided by this principle, we designed a series of voids in which both the radius and depth increase proportionally while keeping d/r constant. This geometrically self-similar scaling enables us to verify whether the resonance wavelength shifts linearly with cavity dimensions, as predicted by theory. We also note the appearance of several high-order modes near a radius of approximately 980 nm. These higher-order resonances exhibit very high Q factors due to delicate interference conditions but are extremely sensitive to structural deviations, making them difficult to realize experimentally or employ in practical devices.

The reflection spectra of this first set of structures are summarized in Fig. 2b. Experimental data agree closely with simulations, demonstrating that the resonance peak and array colours (Supplementary Fig.2) redshifts systematically with increasing void size. The observed wavelength shift is linear and reproducible, confirming the dominance of geometric parameters in determining resonance positions. Notably, the Mie modes remain clearly defined and spectrally distinct throughout the entire scaling process, indicating high structural tolerance and robust modal behaviour.

To further investigate how the spatial distribution of cavity modes evolves with depth, we designed a second set of experiments. This time, the radius was fixed at 865 nm, while the cavity depth was systematically varied from 900 nm to 720 nm. This approach follows the modal evolution path illustrated in Fig. 1c. As shown in Fig. 2c, resonance peaks persist across all depth values, confirming that the modal structure remains intact even as the geometry deviates from the optimal configuration.

As the cavity depth is systematically tuned away from the optimal geometric configuration, the resonance modes not only persist but exhibit continuous evolution in both spectral position and spatial distribution. This behaviour reflects the strong geometric sensitivity and structural tolerance of Mie resonances, which are governed by the total optical path length and internal reflection phase at the dielectric–air interface. As the cavity becomes deeper, the resonant mode migrates downward into the structure, extending the effective optical path and producing a redshift. Conversely, shallower cavities cause the mode to shift upward, closer to the surface, resulting in a slight blueshift. Despite deviation from the optimal enhancement condition, the modal profiles remain well-defined and spectrally distinct, confirming the robust nature of the cavity modes under non-ideal fabrication conditions[32,33].

We emphasize that the two sets of experiments demonstrate the high tunability and modal stability of $Bi_2Te_3$ Mie voids. The resonance wavelength can be linearly adjusted via geometric parameters, while the spatial mode profile remains well-defined and predictable across variations.

**Enhanced photoluminescence from monolayer $WS_2$ via resonantly tuned Mie voids**

To experimentally validate the photoluminescence (PL) enhancement enabled by Mie resonances, we fabricate a set of $Bi_2Te_3$ Mie voids with the same void radius but systematically varied cavity depths. The voids were laterally separated by sufficient distances to avoid optical coupling and were arranged in a regular grid for convenient characterization. Monolayer $WS_2$ was then transferred across the entire patterned area, following the same geometry used in the theoretical simulations (Fig. 1b). The voids were created by discrete ion implantation, allowing us to span both resonant and off-resonant regimes with precise depth control. An optical image of a representative sample is shown in Fig. 3a, where the central region contains the Mie void array while the outer regions remain flat. The $WS_2$ monolayer was transferred continuously across the whole field, ensuring uniform material quality and identical excitation conditions. This configuration allows direct comparison of PL from $WS_2$ positioned over resonant voids, non-resonant voids, and unstructured $Bi_2Te_3$ regions.

We first mapped the spatial PL distribution under broadband white-light excitation (Fig. 3b). A clear enhancement is observed in the void array region compared to the adjacent flat substrate. Since the $WS_2$ layer is continuous, the enhancement originates from the underlying nanostructure rather than material differences. However, due to the low excitation power of the white-light source, PL contrast across different void depths remains weak, making it difficult to resolve the depth dependence predicted in simulation.

To quantify this effect, we performed pointwise PL measurements under 532 nm laser excitation on individual voids within a Mie-void array fabricated with stepped ion doses, corresponding to a depth interval of approximately 30 nm between adjacent voids. The void radius was fixed at 850 nm, consistent with the simulated optimal condition shown in Fig. 1b, while the depth varied from 570 nm to 1020 nm, spanning across the designed resonant depth. Six representative voids were selected, covering a wide range of depths from far-off-resonance to near-resonance. As shown in Fig. 3c, the maximum PL intensity was observed at a cavity depth around 810 nm — closely matching the simulated resonance condition (790 nm). This validates that PL enhancement is governed by the near-field Mie resonance rather than by intrinsic variations in the $WS_2$ film.

We further extracted the relative PL enhancement factors for all measured voids (Fig. 3d). The reference point was chosen as the most off-resonant structure, which did not support a cavity mode and thus exhibited minimal enhancement. Variations in the PL intensity can be interpreted within the standard framework in which the observed signal is jointly governed by the excitation efficiency ($\eta_{ext}$), the internal radiative efficiency ($\eta_{int}$), and the optical out-coupling efficiency ($\eta_{out}$)[34,35]:

$$I_{PL} \propto \eta_{ext} \eta_{int} \eta_{out} \tag{6}$$

Numerical simulations reveal no appreciable enhancement at the excitation wavelength. At 532 nm, the simulated electric-field distribution (Supplementary Fig. 7) shows neither a resonance nor any meaningful local-field amplification for any cavity depth, indicating that the absorbed pump power remains essentially constant across the array.

This conclusion is further supported by pump wavelength dependent PL measurements performed at 490 nm, 520 nm, and 570 nm (Supplementary Fig. 8). In all cases, the PL-enhancement maximum remains pinned to the same cavity depth corresponding to the designed emission resonance near 625 nm, with no measurable shift or systematic dependence on the pump wavelength. These observations conclusively exclude pump-resonant absorption as the dominant mechanism underlying the enhancement. The enhancement reported in Fig. 3d is quantified relative to the least resonant void (depth 570 nm). A comparison with the monolayer yields an even larger apparent enhancement (Supplementary Fig. 6), which can be attributed to the removal of the underlying substrate[36].

Taken together, these results indicate that the enhancement originates predominantly from increased optical out-coupling efficiency enabled by the emission-resonant cavity mode, together with an increase in the local density of optical states arising from the resonant electric-field enhancement at the emission wavelength[34,37]. Further enhancement could, in principle, be achieved by tailoring the Mie-void geometry such that both the pump and emission wavelengths coincide with cavity resonances, providing an additional design parameter for optimising the overall PL efficiency.

**Nonlinear optical response enabled by resonant Mie voids**

Building on the established resonance platform, we investigated how the cavity geometry modulates SHG in the system. A $Bi_2Te_3$ Mie void array was fabricated with a fixed radius of 1190 nm and systematically varied depths cantered around the electric dipole resonance at 870 nm, which was obtained by linearly scaling the 625 nm PL-resonant design while maintaining the same depth-to-radius ratio ($d/r \approx 0.94$), as confirmed by Supplementary Fig. 1. A monolayer $WS_2$ was uniformly transferred on top of the array to ensure consistent material quality and optical excitation across the sample. Under 870 nm fixed-wavelength excitation, the SHG signal was collected from the $WS_2$-covered voids and normalized to the minimum response, corresponding to a non-resonant void depth of 1415 nm. As shown in Fig. 4a, the enhancement factor peaks at ~25× near the designed resonance condition, in agreement with simulations predicting maximum field confinement at a depth of 1121 nm. These results confirm that tuning the void depth offers an effective mechanism for modulating nonlinear optical emission.

SHG, with its quadratic dependence on local electric field intensity, serves as a sensitive probe of resonant enhancement effects. Compared to linear reflection or PL measurements, SHG provides stronger contrast in detecting spatial variations of the optical mode profile. In this system, far-field SHG imaging functions not only as a readout of nonlinear output, but also as an indirect and high-resolution visualization of localized resonant fields at the $WS_2$ interface — without the need for near-field or interferometric methods.

We then assessed the spectral selectivity of the nonlinear response by measuring SHG signals from the $WS_2$ monolayer under pump wavelengths ranging from 850 to 1040 nm in 10 nm steps. As shown in Fig. 4b, the SHG intensity exhibits a clear peak at 870 nm and declines rapidly on either side, consistent with the spectral signature of a localized Mie resonance. This confirms that the nonlinear enhancement is spectrally governed by the geometrically tuned optical mode.

By systematically tuning the pump wavelength from 820 to 1010 nm, we observed that the SHG emission hotspot gradually shifts toward deeper voids in the array. This trend mirrors the redshift in the resonant condition and directly reflects the modal dispersion engineered through cavity geometry. Each excitation wavelength selectively activates cavities whose depths support localized dipolar modes at corresponding frequency. These modes shape the spatial profile of the second-order polarization and govern the efficiency and directionality of the SHG output. The wavelength-dependent migration of the SHG hotspot thus confirms the geometric origin of resonance tuning and highlights the spatially programmable nature of nonlinear emission in this system.

To pinpoint the physical origin of the SHG enhancement, we compared representative voids across the array by correlating experimental SHG images with full-wave electric field simulations. As shown in Fig. 4e, the strongest SHG arises from voids supporting highly symmetric, dipole-like mode profiles. Although the simulated and measured SHG field distributions exhibit minor differences, these can be attributed to the diffraction-limited resolution of the objective lens used in the experiment, which smooths out fine spatial details. In contrast, weaker responses originate from delocalized or higher-order multipolar fields. These observations align with the modal hierarchy established in our theoretical framework, where the dipole mode dominates due to its strong confinement within the air core and efficient spatial overlap with the $WS_2$ monolayer. Off-resonant voids supporting quadrupolar or leaky modes contribute less to SHG due to their weaker confinement. This interpretation is reinforced by the wavelength-resolved SHG mapping in Supplementary Fig. 9, where the observed hotspot follows the simulated modal dispersion shown in Fig. 1d.

## Conclusions

We have developed a novel hybrid photonic platform by stacking a monolayer $WS_2$ on a Mie-void metasurface fabricated in a high-index $Bi_2Te_3$ van der Waals bulk material. Leveraging strong Fresnel reflection at air–dielectric interfaces, the system supports surface-localized electric dipole resonances that significantly enhance the strength of local fields. We have observed experimentally enhanced photoluminescence and second-harmonic generation signals from the $WS_2$ monolayer, with the enhancement rates of approximately 20× and 25×, respectively. By tuning the cavity geometry, we have achieved dual control over the resonance both spectrally and spatially and directly visualized the depth-dependent evolution of the resonant modes through the SHG imaging. Unlike conventional embedded or periodic architectures, this platform enables free-standing integration of 2D materials onto resonators, thereby introducing a new degree of geometrical tunability that can be directly applied to improve previous designs based on bare monolayers[38-41].

Beyond enhanced light-matter interaction, the system serves as a hyperbolic photonic platform for probing and validating the spatial characteristics of individual optical resonances[10,42,43]. The programmable mode geometry, together with far-field imaging capability, enables experimental access to mode migration[44], coupling effects[31,45], and cavity–exciton interactions[40,41] in real space. This opens promising opportunities for nonlinear holography[26], frequency conversion[46], biosensing[5,47], and reconfigurable photonic circuits based on atomically thin materials[48].

## Materials and Methods

Device fabrication and characterization: $Bi_2Te_3$ flakes were mechanically exfoliated from bulk single crystals (HQ Graphene) using adhesive tape and manually transferred onto a rigid copper-tape-coated substrate to ensure mechanical stability and electrical isolation. Circular holes were directly milled into the $Bi_2Te_3$ surface using a focused ion beam (FIB, FEI Helios NanoLab) without any lithographic masking or resist layer. The FIB process was performed at an accelerating voltage of 30 kV with beam currents ranging from 9.7 to 15 pA. Dwell times of 5–15 μs and multi-pass patterning were used to control the final depth of the Mie voids. The geometry of each void, including diameter and depth, was verified by scanning electron microscopy (SEM) using the same FEI Helios NanoLab system.

Monolayer $WS_2$ was obtained via mechanical exfoliation from bulk crystals (2D Semiconductors Inc.) and identified by optical microscopy and photoluminescence (PL) spectroscopy. Selected monolayer flakes were transferred onto the pre-patterned $Bi_2Te_3$ substrates using a dry-transfer method under ambient conditions. After alignment and contact, the sample was heated to 120 °C and maintained for 60 minutes to improve adhesion, followed by slow removal of the transfer stamp.

**Numerical simulations:** For numerical simulations of the eigenmode and scattering spectra, we used the finite-element eigenmode solver and the frequency-domain solver in COMSOL Multiphysics, respectively. In the scattering simulations, the excitation source was a plane wave propagating along the Z-axis with X-oriented polarization, consistent with the experimental configuration. The calculations were performed for an individual void of a specific size, either embedded in a homogeneous medium or placed on a semi-infinite substrate, both surrounded by perfectly matched layers (PMLs) to emulate open boundary conditions.

The geometry of the PML was set to be semi-spherical in the upper domain to match the spherical symmetry of the far-field radiation pattern of a single scatterer, while a cylindrical PML was used in the lower semi-space. This configuration preserves the rotational symmetry of the structure and significantly reduces the total number of finite elements. The fields are strongly confined within the high-index substrate and do not radiate into the far-field in the downward direction, making this approach both accurate and computationally efficient.

The quality factor (Q) of the modes was calculated from the complex resonance frequency, defined as the ratio of the real part to twice the magnitude of the imaginary part. For mode spectrum analysis and parameter optimization, a 2D model with axial symmetry was employed, following the procedure described in Gladyshev's work[49]. For visualization of local fields at the pump and second-harmonic frequencies, a full-wave 3D model was used.

**Linear reflection spectroscopy**: Linear reflectance spectra were acquired using a hyperspectral imaging system (Photonects) integrated with a CCD-based silicon camera (400–1000 nm detection range). Samples were illuminated by a broadband white light source

(Olympus TH4-200), and reflected light was collected through a 5× objective lens under normal incidence. The measurements were performed in the 400–1000 nm spectral range with a 1 nm step size and an integration time of 1 s per frame. Spatially resolved hyperspectral reflection maps were generated by raster scanning across the cavity array, allowing direct correlation between resonance features and void geometry.

**SHG imaging:** Second-harmonic generation (SHG) measurements were carried out using a Zeiss LSM 780 confocal microscope integrated with a tunable Ti: sapphire femtosecond laser (pulse width ~150 fs, repetition rate 80 MHz) as the excitation source. The sample was excited using a 50× objective lens (NA = 0.85), and SHG signals were collected in both reflection and transmission geometries. For fixed-wavelength experiments, the fundamental laser wavelength was set to 900 nm; for wavelength-dependent measurements, the excitation wavelength was tuned across 820–1040 nm in 20 nm increments. All SHG signals were filtered using short-pass or bandpass filters to isolate the frequency-doubled component and recorded using a photon-counting detector integrated into the system as detailed in Supplementary Fig.5. The excitation power was actively stabilized at 80 μW throughout all measurements, and the fluctuation was confirmed to remain below 1 μW (≈ 1%) during the entire scanning process, ensuring the reliability of the observed SHG enhancement.

**PL Spectroscopy:** PL measurements are performed using a Horiba LabRAM system equipped with a confocal microscope, a silicon-based charge-coupled device (CCD) detector (400–1000 nm), and a 532 nm diode-pumped solid-state laser as the excitation source. The excitation beam was focused onto the sample surface using a 50× objective lens (numerical aperture (NA) = 0.55), producing a near-Gaussian spot with a diameter of approximately 0.5 μm. All PL signals were collected in a back-reflection configuration under ambient conditions. The laser spot was precisely positioned on the monolayer $WS_2$ region above individual Mie voids to probe the localized emission response of each cavity.

## Acknowledgements

Y.K. thanks Mario Hentschel for useful comments and suggestions. This work was supported by the Australian Research Council (Grant No.: DP240101011, DP220102219, DP210101292) and the National Health and Medical Research Council (NHMRC; ID: GA275784). The authors acknowledge the use of the fabrication facilities at the ACT Node of the NCRIS-enabled Australian National Fabrication Facility at the Australian National University (ANFF-ACT). K.K. acknowledges support from the Australian Research Council through the Discovery Early Career Researcher Award (Grant No.: DE250100419).

**Competing interests** The authors declare that they have no competing financial interests or any other conflict of interest

## Author Contributions

Y. L., Y.K., and K.K. conceived the idea and supervised the project; Z.L. prepared the samples; K.K. performed numerical simulations. Z.L., Z.C, P.T. carried out the optical measurements; Z.L. and Y.L. analysed the data; Z.L. and Y.L. prepared the initial draft, and all authors contributed to the preparation of the manuscript.

## Supplementary Information

All additional data, information, and methods are presented in the supporting information file available online. The figures and information in the supporting information have been cited at appropriate places in this manuscript.

a 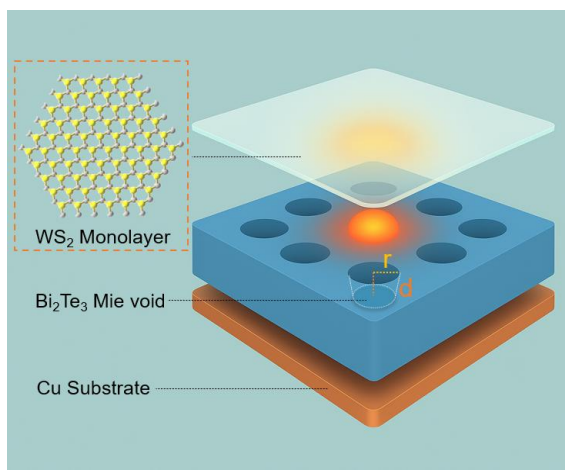

b 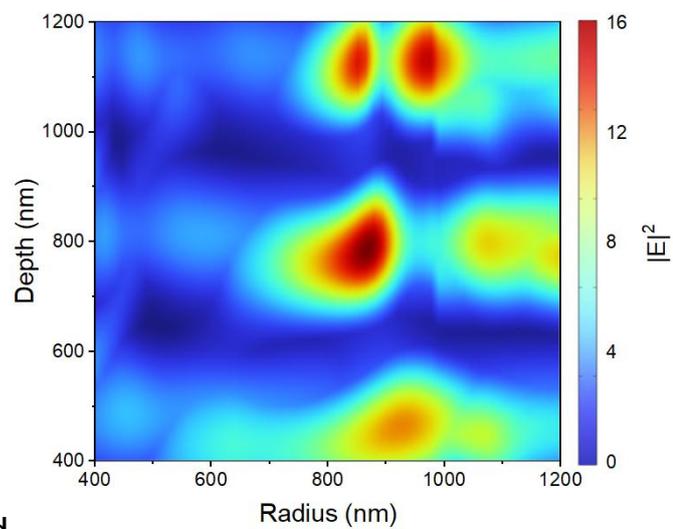

c 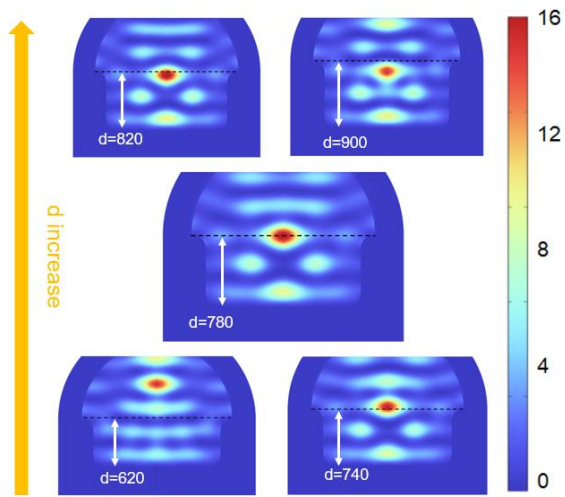

d 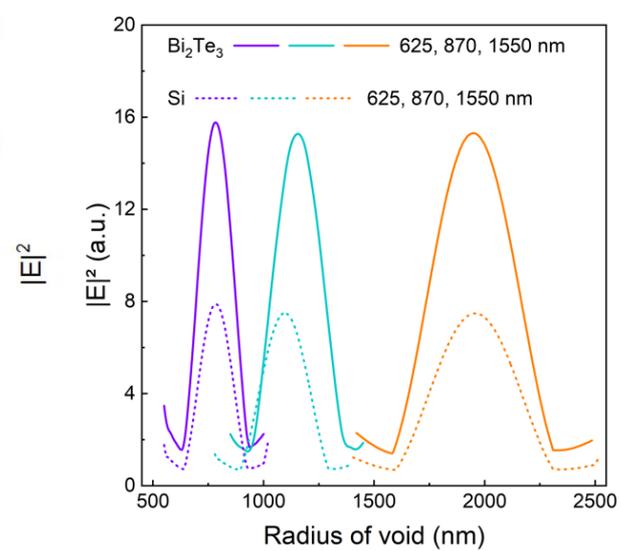

Figure 1

a

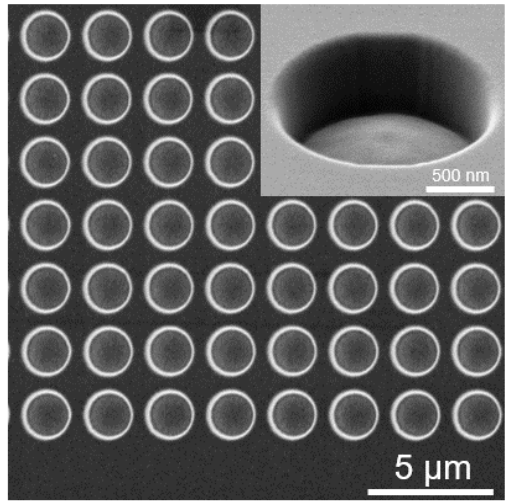

b

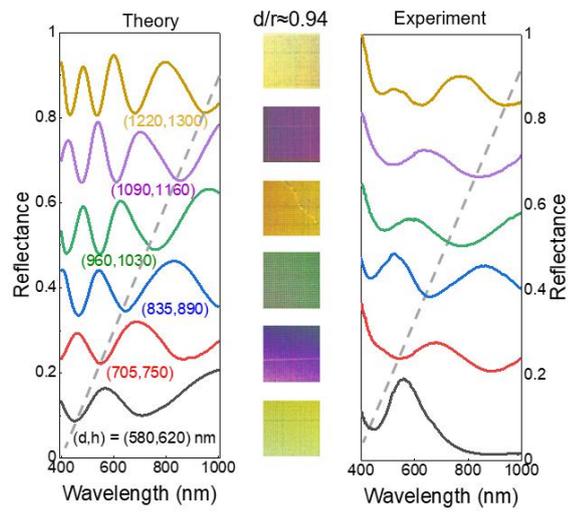

c

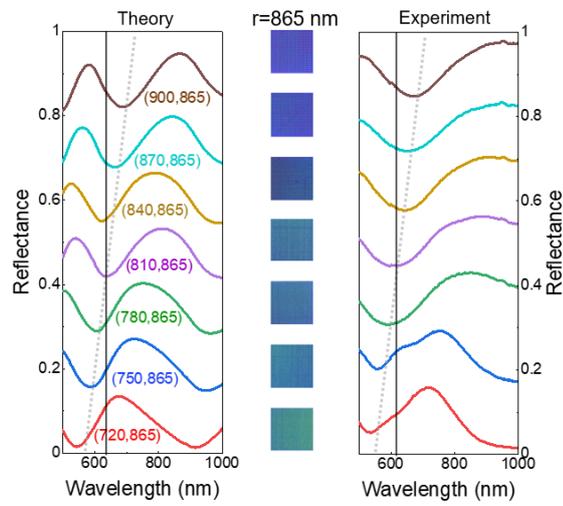

Figure 2

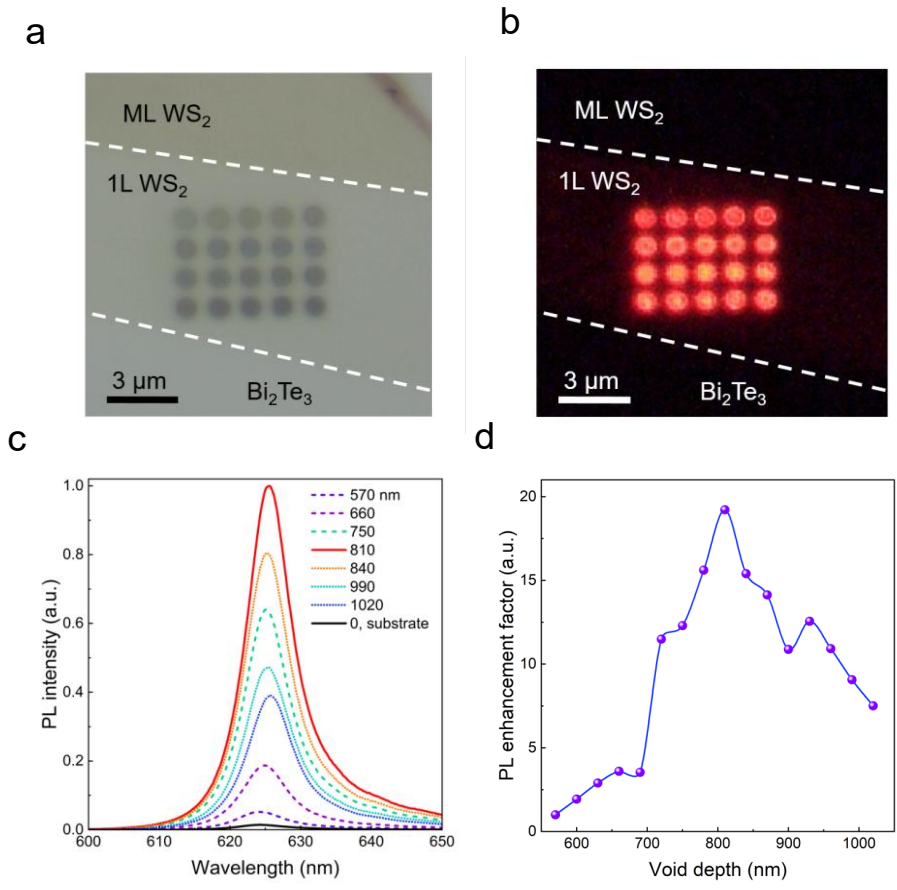

Figure 3

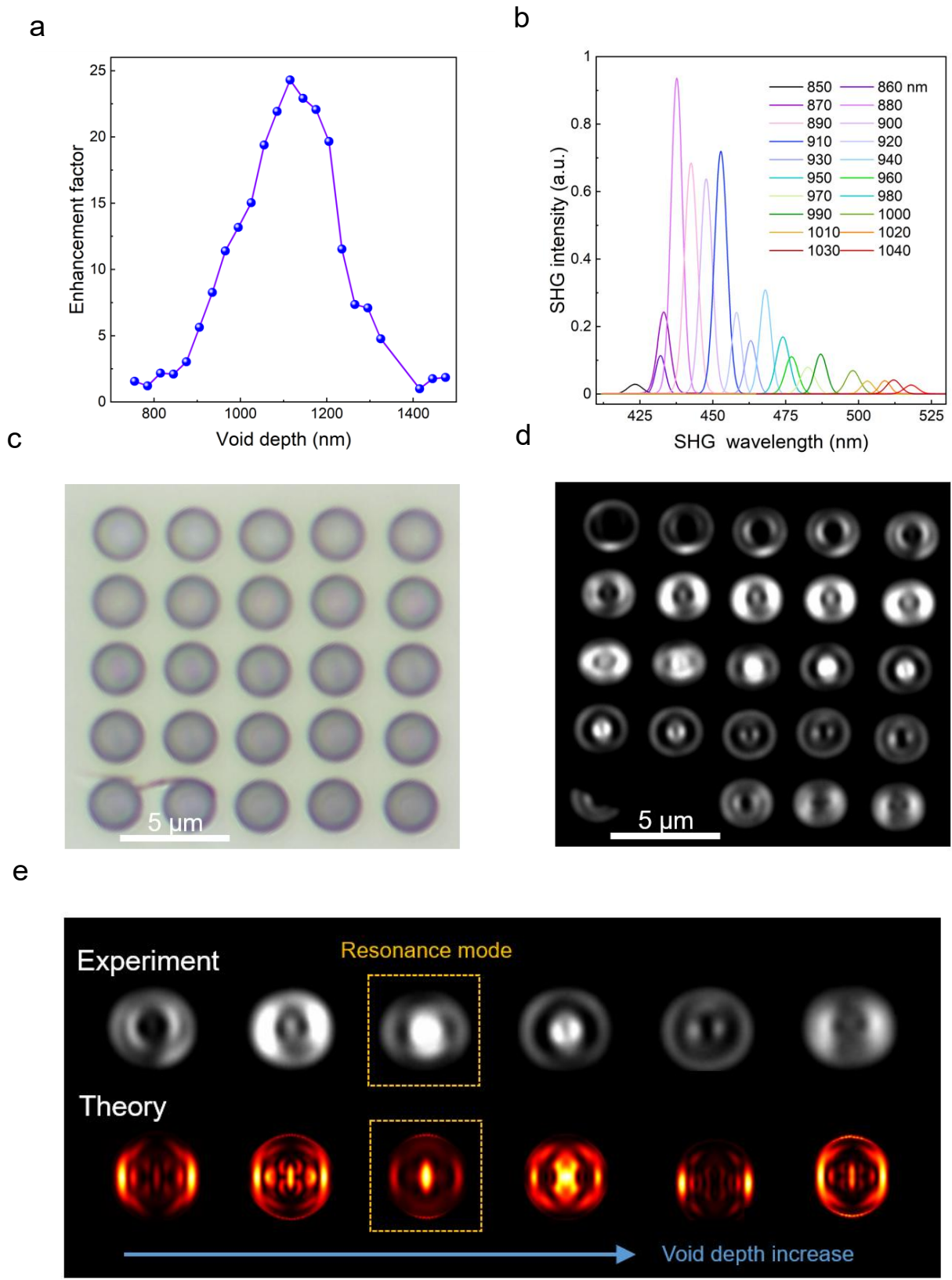

Figure 4

# Figure Captions

**Figure 1 | Concept of Mie-void heterostructures. a**, Schematic illustration of a monolayer $WS_2$ stacked on a $Bi_2Te_3$ Mie void, forming a fully two-dimensional van der Waals heterostructure platform. The monolayer $WS_2$ is shown as an ultrathin transparent layer, with the localized Mie resonance field excited at the central cavity and projecting a corresponding highlighted region onto the overlying monolayer. The molecular structure of $WS_2$ is highlighted in the dashed orange box on the left. The parameters r and d denote the void radius and depth, respectively. **b**, Simulated in-plane electric field enhancement ($|E|^2$) evaluated at the top surface of the void structure, corresponding to the plane of the $WS_2$ monolayer, plotted as a function of void radius and depth under 625 nm, corresponding to the A exciton resonance of $WS_2$. The results reveal that the maximum field enhancement occurs at a void radius of approximately 850 nm and a depth of 780 nm, aligning with the fundamental Mie resonance mode. **c**, Simulated side-view electric field distribution for Mie void with the resonances of 625 nm, illustrating the spatial positioning of the resonance mode as a function of void depth. When the void depth is 850 nm (consistent with panel b), the Mie resonance is localized near the void surface, maximizing interaction with the $WS_2$ monolayer. As the void depth increases, the resonance gradually shifts downward. **d**, Comparative simulations of electric field enhancement $|E|^2$ for $Bi_2Te_3$ and Si Mie voids as a function of void radius, with the depth fixed at the resonance condition for each wavelength. Results are shown for three resonance wavelengths (625 nm, 780 nm, and 1550 nm).

**Figure 2 | Resonant characteristics of Mie voids fabricated in $Bi_2Te_3$. a**, Top-view scanning electron microscopy (SEM) image of a periodic $Bi_2Te_3$ Mie void. Inset: Tilted-view SEM image of a single void. **b**, Comparison of simulated and measured resonance spectra for Mie voids with varying radius and depth and fix the ratio depth-to-radius to be 0.94 , which remain in the on-resonance regime across the parameter space. The left panel shows simulated reflectance spectra as a function of both radius and depth, illustrating the broad tunability of the Mie resonance across 450–950 nm. The middle panel presents an optical microscope image of the fabricated void array from Supplementary Fig.4. The right panel shows experimentally measured reflectance spectra corresponding to the array. **c**, Comparison of simulated and measured reflectance spectra for Mie voids with depths ranging from 720 nm to 900 nm in 30 nm increments, corresponding to both on-resonance and off-resonance conditions, with the radius fixed at 865 nm. The left panel shows simulated spectra highlighting the depth-

dependent resonance shifts. The middle panel displays an optical microscope image of the array. The right panel presents the measured reflectance spectra.

**Figure 3 | Photoluminescence enhancement of monolayer WS$_2$ via Bi$_2$Te$_3$ Mie voids. a**, Optical microscope image of monolayer WS$_2$ transferred onto patterned Bi$_2$Te$_3$ substrates containing Mie void arrays. Two distinct regions are identified: WS$_2$ stacked on flat Bi$_2$Te$_3$ (left) and WS$_2$ stacked on Mie voids (right). The void arrays are engineered with varying depths to enable resonance tuning. **b**, PL Mapping image corresponding to the same region as in (a). The PL intensity is clearly enhanced in the WS$_2$ region located above the Mie voids, while it remains much weaker on the flat Bi$_2$Te$_3$ region, directly confirming the critical role of Mie resonance in emission enhancement. The spatial variation in PL reflects local enhancement from voids of different depths, which serves as the basis for the pointwise analysis in panels c and d. **c**, Pointwise PL measurements were performed under 532 nm laser excitation across a 4×4 Mie void array with systematically varied depths. Six representative voids were selected, covering both resonant and off-resonant conditions with ~30 nm depth intervals. The strongest PL signal occurs at the void depth of ~810 nm, consistent with the simulated resonance. All other depths yield progressively weaker emission. **d**, Relative PL enhancement factors as a function of void depth. The most off-resonant void is used as the reference (enhancement = 1). The Mie-resonant void yields an additional ~20× enhancement. This effect is complementary to the enhancement already present in suspended or cavity-integrated WS$_2$. Supplementary Fig.6 further present comparisons with Bi$_2$Te$_3$ and other substrates.

**Figure 4 | Depth-dependent SHG response and field distribution in WS$_2$/Bi$_2$Te$_3$ Mie-void heterostructures a,** Relative SHG enhancement factors as a function of void depth, extracted from a 5×5 Mie void array with a fixed radius of 1190 nm, designed to support resonance at 870 nm. The void with a depth corresponding to a simulated non-resonant condition at 1415 nm is used as the normalization reference (enhancement = 1). The maximum enhancement (~25×) is observed at a void depth resonant under 1115 nm excitation, in close agreement with the simulated resonance at 1121 nm. This enhancement reflects only the contribution from resonance tuning and does not account for other cavity or substrate effects**. b,** Wavelength-dependent SHG intensity measured from a Mie void with fixed geometry supporting resonance at 870 nm. The SHG signal peaks at 870 nm pump (435 nm SHG) and decreases progressively as the pump wavelength is detuned. The excitation wavelength is scanned from 830 nm to 1040 nm in 10 nm increments. **c,** Optical microscope image of a monolayer WS$_2$ flake stacked on a

Bi$_2$Te$_3$ Mie void array with systematically varying void depths. The depth increases from the top-left to the bottom-right of the array, with the central void designed to be on resonance. The WS$_2$ uniformly covers the entire patterned region, enabling direct comparison between voids of different geometries. **d,** SHG intensity map of the region shown in (c), measured under 870 nm excitation. The SHG signal is maximized at the center of the array and decreases for shallower or deeper voids along the diagonal depth gradient. **e,** Comparison between experimental SHG mapping and simulated electric field distributions for five representative voids selected from the array in Fig. 4d. For each void, the top panel shows a magnified SHG intensity map under 870 nm excitation, and the bottom panel presents the corresponding simulated $|E|^2$ distribution at the same wavelength. The voids supporting strong SHG signals exhibit electric dipole-like field confinement, while non-resonant voids show more delocalized or multipolar field patterns.